\documentclass[12pt]{article}
\usepackage{graphicx}
\usepackage{amssymb}
\newcommand{\beq}{\begin{equation}}
\newcommand{\eeq}{\end{equation}}
\newcommand{\ep}{\mbox{${\varepsilon}$}}

\begin{document}

\begin{center}
{\Large \bf  Electric dipole moments, from $e$ to $\tau$}\\[3mm]
A.G.~Grozin\footnote{A.G.Grozin@inp.nsk.su},
I.B.~Khriplovich\footnote{khriplovich@inp.nsk.su},
and A.S.~Rudenko\footnote{saber\_@inbox.ru}\\
Budker Institute of Nuclear Physics,
630090 Novosibirsk, Russia,\\
and Novosibirsk University
\end{center}

\begin{abstract}
We derive an upper limit on the electric dipole moment (EDM) of the $\tau$-lepton,
which follows from the precision measurements of the electron EDM.
\end{abstract}

\vspace{3mm}

The best upper limit on the electric dipole moment of electron, as derived in a spectroscopic,
almost table-top experiment, is \cite{bcr}
\beq\label{e}
d_e/e = (0.7\pm 0.7)\times 10^{-27}\; {\rm cm}.
\eeq
The best result for the muon EDM, following from the
measurements at the dedicated muon storage ring, is \cite{jmb}
\beq\label{mu}
d_\mu/e = (0.3\pm 0.3)\times 10^{-18}\; {\rm cm}.
\eeq

We investigate here the $\tau$-lepton EDM. The information on it was obtained up to now
from the analysis of the process $e^+e^- \to \tau^+\tau^-$. The
regular mechanism of this reaction is due to the intermediate
photon and $Z$-boson with the usual electromagnetic and
neutral-current $\tau$ couplings, $\gamma\tau^+\tau^-$ or
$Z\tau^+\tau^-$, correspondingly. In both cases $\tau$-leptons are produced in the
triplet states, $^3S_1$ and $^3D_1$ for the vector couplings, and
$^3P_1$ for the axial one. Meanwhile, when produced via CP-odd EDM
vertex, the $\tau$-leptons are in the singlet $^1P_1$ state.
Obviously, the two amplitudes do not interfere in the
cross-section for unpolarized $\tau$-leptons. They do interfere
however if the polarizations of final $\tau$ are measured. In this
way, the following result for the EDM of $\tau$-leptons was
obtained in $e^+e^-$ collisions at the center-of-mass energy $2E \simeq
10$~GeV~\cite{Inami}:
\beq\label{in}
d_\tau/e = (1.15\pm 1.70)\times 10^{-17}\; {\rm cm}.
\eeq
On the other hand, the CP-odd amplitude due to the EDM vertex,
being  squared, contributes of course to the total annihilation
cross-section $e^+e^- \to \tau^+\tau^-$. Moreover, this $^1P_1$
contribution to the differential cross-section behaves as
$\sin^2\theta$, as distinct from the regular $1 + \cos^2\theta$
(in the ultrarelativistic limit) for the regular triplet channels.
The corresponding analysis of the differential
cross-sections at $2E \simeq 35$ GeV resulted in the following upper limit on the
dipole moments \cite{del Aguila}:
\beq\label{da}
d_\tau/e< 1.4 \cdot 10^{-16}\; {\rm cm}
\eeq

Better upper limit
\beq\label{es}
d_\tau/e <1.1 \cdot 10^{-17}\; {\rm cm}
\eeq
was obtained from the partial width $\Gamma(Z \to
\tau^+ \tau^-)$ measured at LEP in Z peak \cite{Escribano}.
However, it used model-dependent relationship between the weak and
electric dipole moments.

At last, quite recent analysis \cite{br} of the LEP-II data for the total  annihilation
cross-section $e^+e^- \to \tau^+\tau^-$ at \ $2E \simeq 200$ GeV results in the upper limit
\beq\label{br}
d_\tau/e <3 \cdot 10^{-17}\; {\rm cm}.
\eeq

\begin{figure}[ht]
\begin{center}
\begin{picture}(168,38)
\put(19,21){\makebox(0,0){\includegraphics{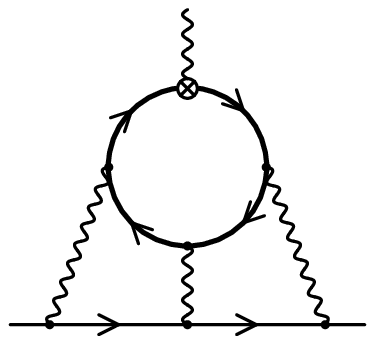}}}
\put(19,0){\makebox(0,0)[b]{a}}
\put(63,21){\makebox(0,0){\includegraphics{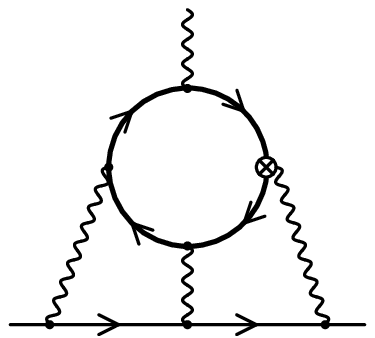}}}
\put(63,0){\makebox(0,0)[b]{b}}
\put(107,21){\makebox(0,0){\includegraphics{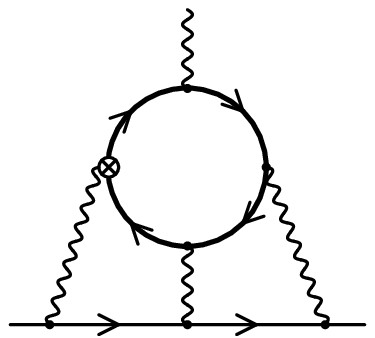}}}
\put(107,0){\makebox(0,0)[b]{c}}
\put(149,21){\makebox(0,0){\includegraphics{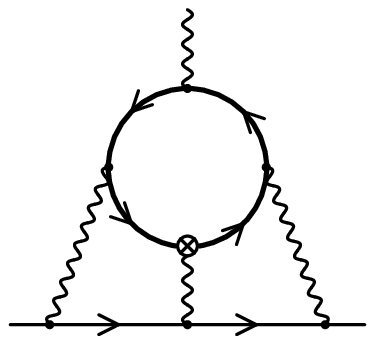}}}
\put(149,0){\makebox(0,0)[b]{d}}
\end{picture}
\end{center}
\caption{The $\tau$ electric dipole moment contribution to the electron EDM}
\label{Fig}
\end{figure}

We will calculate the contribution of the $\tau$ lepton EDM into
the electron electric dipole moment. This contribution is
described by the diagrams of the type presented in Fig.~\ref{Fig}.
Here the loop is formed by the $\tau$ line, and the lower solid
line is the electron one. The upper wavy line corresponds to the
external electric field. The crossed vertices refer to the
electromagnetic interaction of the $\tau$ EDM. All 6 permutations of the vertices along the electron line
should be considered. The contributions b and c are equal.

The discussed effect is similar to the contribution of light-by-light scattering via muon loop
to the electron magnetic moment~\cite{rem}.
The general structure of the resulting contribution to the
electron EDM is rather obvious:
\beq\label{est}
\Delta d_e = a \frac{m_e}{m_\tau} \left(\frac{\alpha}{\pi}\right)^3 d_\tau\,,
\eeq
where $a$ is some numerical factor (hopefully, on the order of
unity). The factor $m_e$ originates from the necessary helicity-flip
on the electron line. Then, $1/m_\tau$ is dictated by the obvious dimensional
argument.

The diagram of Fig.~\ref{Fig}a is just the matrix element
${<}e|\bar{\tau}\sigma^{\mu\nu}\tau|e{>}=C\bar{u}\sigma^{\mu\nu}u$.
We use dimensional regularization with $d=4-2\ep$ dimensions
and the method of regions (see the textbook~\cite{Smirnov}).
Only the region where all 3 loops are hard
(loop momenta $\sim m_\tau$) contributes
to the leading power term~(\ref{est});
therefore, there are no logarithms $\ln(m_\tau/m_e)$
(contributions of regions with 1 or 2 hard loops are suppressed
by an extra factor $(m_e/m_\tau)^2$).
In this hard region, the problem reduces to 3-loop vacuum integrals
with a single mass $m_\tau$ belonging
to the simpler topology $B_M$~\cite{Broadhurst}.
We perform the calculation in arbitrary covariant gauge,
and use the \textsc{Reduce} package \textsc{Recursor}~\cite{Broadhurst}
to reduce scalar integrals to 2 master integrals.
Gauge-dependent terms cancel, and we get
\begin{equation}
C = \frac{m_e}{m_\tau} \frac{e^6 m_\tau^{-6\ep}}{(4\pi)^{3d/2}}
\Gamma^3(\ep)
\frac{8}{d(d-1)(d-5)} \left[ - 2 \frac{2d^2-21d+61}{d-5}
+ \frac{d^4-9d^3+8d^2+84d-126}{2d-9} R \right]\,,
\label{C}
\end{equation}
where
\begin{equation}
R =
\frac{\Gamma(1-\ep) \Gamma^2(1+2\ep) \Gamma(1+3\ep)}%
{\Gamma^2(1+\ep) \Gamma(1+4\ep)} =
1 + 8 \zeta(3) \ep^3 + \cdots\,,
\label{R}
\end{equation}
and $\zeta$ is the Riemann $\zeta$-function.
All divergences cancel, and we arrive at a finite contribution to $a$:
\beq\label{1}
a_1 = \frac{3}{2}\,\zeta(3) - \frac{19}{12}
\eeq

In order to calculate the contribution of Fig.~\ref{Fig}b, c, d,
we expand the corresponding initial expressions in the external photon momentum $q$ up to the linear term.
The EDM vertex contains $\ep^{\mu\nu\alpha\beta}$;
we put this factor aside, and calculate tensor diagrams with 4 indices.
After summing all diagrams, the result is finite;
now we can set $\ep\to0$,
and multiply by $\ep^{\mu\nu\alpha\beta}$ (cf.~\cite{Larin}).
The result has the structure of a tree diagram with the electron EDM vertex
$\ep^{\mu\nu\alpha\beta}q_\nu\sigma_{\alpha\beta}$.
The gauge-dependent terms in it cancel (exactly in $d$), as well as the
divergences. This contribution to $a$ is
\beq\label{2}
a_2 = \frac{9}{4}\,\zeta(3) - 1\,.
\eeq

As an additional check of our programs, we have reproduced the leading power term in the contribution to the electron magnetic moment originating from the light-by-light scattering via the muon loop (formula~(4) in~\cite{rem}).

Our final result for the numerical coefficient is
\beq\label{fin}
a= a_1 + a_2 = \frac{15}{4}\, \zeta(3) - \frac{31}{12} = 1.924\,.
\eeq
With this value of $a$, the discussed contribution to the electron EDM is
\beq
\Delta d_e = 6.9 \times 10^{-12}\, d_\tau.
\eeq
Combining this result with the experimental one~(\ref{e}) for the electron EDM, we arrive at
\beq\label{fin1}
d_\tau/e = (1 \pm 1)\times 10^{-16}\,\mathrm{cm}\,.
\eeq

Let us note however that, strictly speaking, the contributions of diagram 1a and diagrams 1b, c, d
refer to somewhat different kinematical regions. While in the last case all 3 momenta
entering the EDM vertex are hard, on the order of magnitude about $m_\tau \sim 1$ -- 2 GeV, in the first case
the photon entering this vertex is soft, of vanishing momentum. Still, one may expect that the effective EDM interaction is formed at momenta much higher than $m_\tau$, so that this difference is not of much importance. Besides, the contribution of diagram 1a is anyway numerically small. Thus, result (\ref{fin1}) is valid at least
for all momenta about $m_\tau \sim 1$ -- 2 GeV.

On the other hand, the previous results (\ref{in}),
(\ref{da}), (\ref{es}), and (\ref{br}) refer to much larger photon momenta of 10, 35, 90, and 200 Gev,
respectively.

\vspace{5mm}

{\bf Acknowledgements.}
The work was supported in part by the Russian Foundation
for Basic Research through Grant No. 08-02-00960-a.

\end{document}